\documentclass[10pt,conference]{IEEEtran}
\IEEEoverridecommandlockouts
\usepackage{cite}
\usepackage{algorithm}
\usepackage{algorithmic}
\usepackage{amsmath,amssymb,amsfonts}
\usepackage{graphicx}
\usepackage{textcomp}
\usepackage{xcolor}
\def\BibTeX{{\rm B\kern-.05em{\sc i\kern-.025em b}\kern-.08em
    T\kern-.1667em\lower.7ex\hbox{E}\kern-.125emX}}

\makeatletter
\newcommand{\linebreakand}{%
  \end{@IEEEauthorhalign}
  \hfill\mbox{}\par
  \mbox{}\hfill\begin{@IEEEauthorhalign}
}
\makeatother

\begin{document}

\title{CSSAM: Code Search via Attention Matching of Code Semantics and Structures}

\author{\IEEEauthorblockN{1\textsuperscript{st} Yi Hu}
\IEEEauthorblockA{\textit{School of Cyber Science and Engineering} \\
\textit{WuHan University}\\
WuHan, China \\
csehuyi@whu.edu.cn}
\and
\IEEEauthorblockN{2\textsuperscript{nd} Bo Cai}
\IEEEauthorblockA{\textit{School of Cyber Science and Engineering} \\
\textit{WuHan University}\\
WuHan, China \\
caib@whu.edu.cn}
\and
\IEEEauthorblockN{3\textsuperscript{rd} Yao Xiangyu}
\centerline~\IEEEauthorblockA{\textit{School of Cyber Science and Engineering} \\
\textit{WuHan University}\\
WuHan, China \\
yu.yaoxiang@whu.edu.cn}
}

\maketitle

\begin{abstract}
Code search greatly improves developers' coding efficiency by retrieving reusable code segments from open source repositories with natural language queries. Despite the continuous efforts in improving both the effectiveness and efficiency of code search, two issues remained unsolved. First, programming languages have inherent strong structural linkages, and feature mining of code as text form would omit the structural information contained inside it. Second, there is a potential semantic relationship between code and query, it is challenging to align code and text across sequences so that vectors are spatially consistent during similarity matching.

To tackle both issues, in this paper, a code search model named CSSAM (Code Semantics and Structures Attention Matching) is proposed. By introducing semantic and structural matching mechanisms, CSSAM effectively extracts and fuses multi-dimensional code features. Specifically, the cross and residual layer was developed to facilitate high-latitude spatial alignment of code and query at the token level. By leveraging the residual interaction, a matching module is designed to preserve more code semantics and descriptive features, that enhances the adhesion between the code and its corresponding query text. Besides, to improve the model's comprehension of the code's inherent structure, a code representation structure named CSRG (Code Semantic Representation Graph) is proposed for jointly representing abstract syntax tree nodes and the data flow of the codes. According to the experimental results on two publicly available datasets containing 540k and 330k code segments, CSSAM significantly outperforms the baselines in terms of achieving the highest SR@1/5/10, MRR, and NDCG@50 on both datasets respectively. Moreover, the ablation study is conducted to quantitatively measure the impact of each key component of CSSAM on the efficiency and effectiveness of code search, which offers the insights into the improvement of advanced code search solutions.
\end{abstract}

\begin{IEEEkeywords}
code search, graph representation, text matching, attention mechanism, graph attention network
\end{IEEEkeywords}

\section{Introduction}
With the popularity of open source communities, the amount of code hosted on platforms like Github and StackOverflow is increasing day by day, providing more possibilities for code reuse by program developers. In the face of huge code resources, how to accurately find the corresponding code segment according to user intent has become a popular research problem. Code search has gone through two stages, the early ones are based on information retrieval (IR) methods, such as Portfolio\cite{DBLP:conf/icse/McMillanGPXF11} proposed by  McMillan et al. and CodeHow\cite{2015CodeHow} proposed by Fei et al. These IR-based methods only treat code as a text fragment like natural language, and perform natural language to code fragment retrieval work with the help of search engine ideas. However, it is obvious that code snippets and natural language are heterogeneous and there are considerable differences between them, and the proposed deep learning-based code search methods can better solve this problem. In 2016, CODEnn\cite{DBLP:conf/icse/GuZ018} proposed by Gu et al. was the first method to use deep learning to solve the code search problem. It embeds natural language and code snippets into a high-dimensional vector space through a neural network, and then use the cosine distance to judge the similarity between them.
Compared with information retrieval based models, deep learning (DL) based models can capture higher dimensional information, so the results are also much better than IR models.

Like IR models, DL models also have their limitations, even though existing studies have been able to understand and analyze natural language by neural networks, they cannot understand programming language well. In CODEnn\cite{DBLP:conf/icse/GuZ018}, the authors try to use tokens, api sequence, and method name of code segments to extract semantic features of the code, and this approach proves to be useful. In the code2vec\cite{DBLP:journals/pacmpl/AlonZLY19} model, an attempt is made to use the abstract syntax trees(ASTs) as a feature of the code for embedding. Similarly there is the proposed SBT\cite{DBLP:conf/iwpc/HuLXLJ18} traversal of the abstract syntax tree AST. All these models process the AST of the code as sequence features first, and then learn its features using RNN or LSTM. For such topologies as AST, graph neural networks should be able to embed them better in theory. Wan et al. proposed MMAN\cite{DBLP:conf/kbse/WanSSXZ0Y19} to learn the code attribute graphs of AST, CFG, DFG, etc. as features with gated neural networks GGNN\cite{DBLP:journals/corr/LiTBZ15}, which achieved very good results, but the input features are too scattered.

 The success of large-scale natural language models has led to the emergence of pre-training-based code search models, these encoder models (e.g., CoBert\cite{DBLP:conf/emnlp/FengGTDFGS0LJZ20}, CodeT5\cite{DBLP:conf/emnlp/0034WJH21}) are designed to embed the code and natural language descriptions into the same feature space, then the cosine or L2 similarity of these vectors is computed, but this type of encoder models cannot retain the deep code structure and semantic information,  because of its large training size, it is difficult to make its results converge, besides they could not learn the dependencies between variables in the code and the transition process of the program state. Moreover, despite the fact that the aforementioned models can capture the semantic information of individual code fragments or query text, they are hard to investigate the semantic relationships within the code and between the code and the description text at a finer granularity.
\begin{figure*}[ht]
  \centering
  \includegraphics[width=\linewidth]{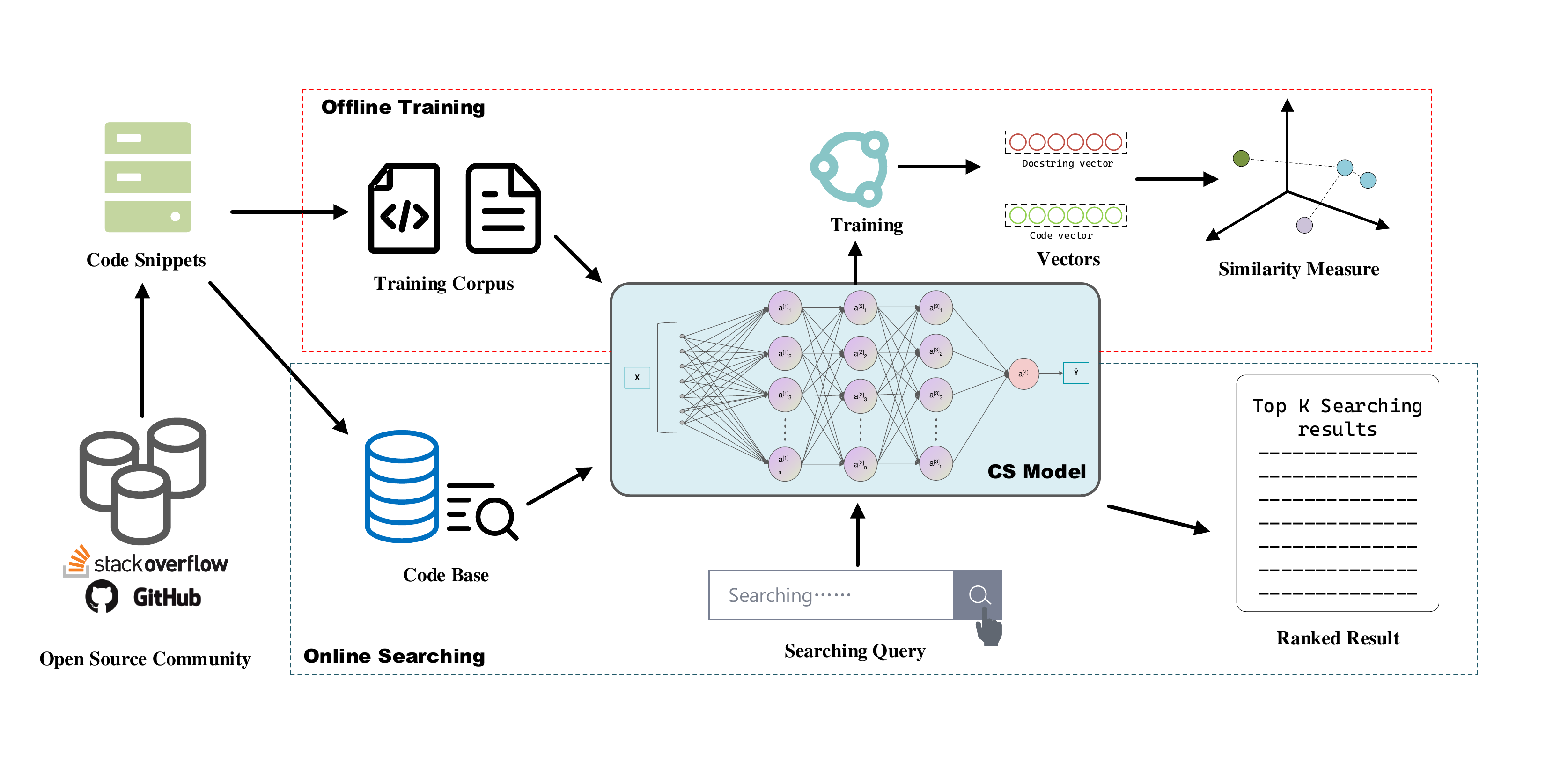}
  \caption{The workflow of our code search model CSSAM. The first thing is to collect code snippets, and then it can be divided into two stages: offline training and online searching. The offline stage uses the processed code-description pairs to train our code search model. After training, it can be used for online code search task.}
\end{figure*}

Based on the existing research, this paper proposes a new structured code search model CSSAM, which is more effective than the existing state-of-the-art model, and the contributions of this paper are as follows:
\begin{itemize} 
\item A new deep learning-based code search model CSSAM is proposed, which semantically extracts and matches codes and descriptions by introducing multiple levels such as semantic level and structural level matching modules.
\item A new code representation structure, the Code Semantic Representation Graph (CSRG) is proposed, which is based on an abstract syntax tree, aggregates nodes, and incorporates data flow features so that the structure contains more semantic information.
\item A code-description matching module CRESS is proposed and used, and a matching module based on residual interaction is designed in this paper, which enhances the description ability of words and phrases by cascading residual information and attention mechanism, thus preserving more textual features of codes and descriptions, while introducing a weight sharing mechanism to match codes and descriptions at the semantic level.
\item The CSSAM proposed in this paper can be accurately trained and tested on two publicly available large datasets, and the experimental results are better than the classical models.
\end{itemize}

The remainder of this paper is organized as follows. Section 2 introduces the related work of this paper. Section
3 presents our proposed model CSSAM. Section 4 describes the experiment setup. Section 5 and Section 6 show the experimental results and discussion respectively. Section 7 concludes the work and presents future works.

\begin{figure*}[htb]
  \centering
  \includegraphics[width=\linewidth]{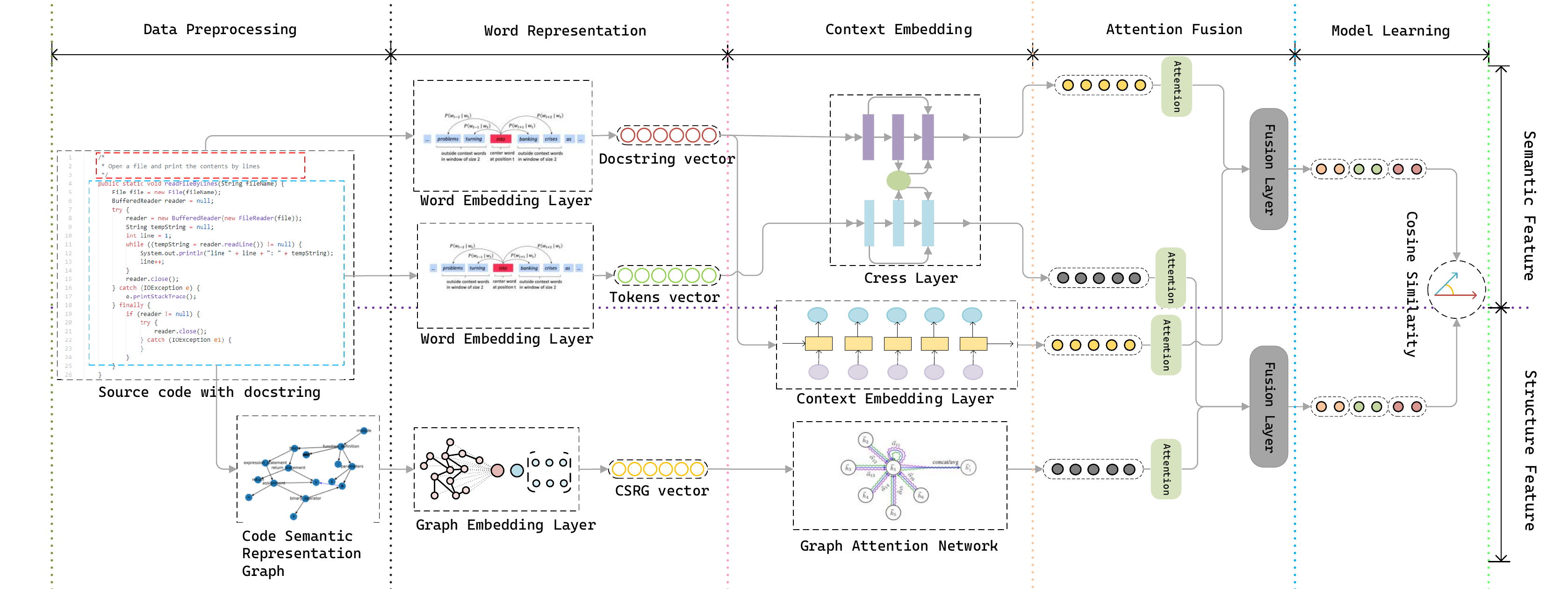}
  \caption{The network architecture of our proposed CSSAM model. We first extract the $<code, description>$ pairs from training corpus. We then parse the code snippets into graph modalities, i.e., tokens, CSRG. Then the training samples are fed into the network as input. (a) Embedding layer. We first learn the representation of each token and node via fasttext, graph embeding. (b) Representation layer. We use LSTM and GAT to learn the representation of tokens and CSRGs. (c) Fusion attention layer. We design an attention layer to assign different weight on different parts for each modality, and then fuse the attended vector into a single vector. (d) Ranking learning. We map the comment description representation and code representation into an intermediate semantic common space and design a ranking loss function to learn their similarities.}
\end{figure*}

\section{Related Work}

In this section, we mainly introduce and discuss the related work of this paper from three aspects: code representation based on deep learning, text matching, attention mechanism and graph neural network.

\subsection{Code Representation}

In early research, code was treated as a natural language, and it was taken for granted to process code fragments using natural language models. Later, Gu et al.'s research and experiments\cite{DBLP:conf/icse/GuZ018} proved that method names and API sequences in code are also crucial for code feature extraction, and these methods were used in code search and code annotation generation. The best representation of programming languages is the abstract syntax tree AST, and in 2016, the distributed code2vec model for learning code using AST paths was proposed, which was used in method name prediction of code fragments with very good results, and the model was further demonstrated to contain a large amount of code semantic information in the AST.

Since code2vec was proposed, people have started to study the topological representation of code. code2vec uses distributed representation of words to learn the vector representation of tokens in code by traversing the AST\cite{codeclone,ARoma} paths. Hu et al.'s proposed SBT\cite{DBLP:conf/iwpc/HuLXLJ18} defines a rule to traverse the AST and convert the topological structure into a sequence structure. Wan et al.’s proposed MMAN\cite{DBLP:conf/kbse/WanSSXZ0Y19} converts the topological structure of code into a sequence structure. MMAN treats the tree structure of the code (ASTs) representation\cite{DBLP:journals/tkdd/LingWWPMXLWJ21,DBLP:conf/iclr/ZugnerKCLG21} as a special graph structure and uses graph neural networks to learn the embedding of codes.

\subsection{Text Matching}
Text matching\cite{bertattention,aclsss,transformermat} is an important fundamental problem in natural language processing and can be applied to a large number of natural language processing (NLP) tasks, such as information retrieval, question and answer systems, paraphrasing problems, dialogue systems, machine translation, etc. These NLP tasks have certain similarities with text matching problems. Similarly, the code search task can be abstracted as a text matching problem, i.e., determining whether the code in the code base matches the query statement. With the successful use of deep learning in the fields of computer vision, speech recognition, and recommender systems, there has been much research in recent years devoted to applying deep neural network models to natural language processing tasks to reduce the cost of feature engineering. The baseline model for deep code search is the classical baseline model based on text matching.

Traditional text matching techniques include the vector space model VSM\cite{DBLP:conf/icml/LeM14,DBLP:journals/jmlr/BengioDVJ03, DBLP:journals/corr/abs-1301-3781}, commonly known as the ``bag of word model", which represents words in a dictionary space generated from text; TF-IDF\cite{DBLP:conf/sigir/SaltonY73}, which adds word frequency weights to VSM; and the BM25\cite{DBLP:conf/sigir/RobertsonW94} algorithm, which calculates the match between web fields and query fields by the degree of coverage of these fields. The higher the score, the better the match between the web page and the query; these algorithms mainly solve the matching problem at the lexical level, or the similarity problem at the lexical level. In fact, matching algorithms based on lexical overlap have great limitations because they ignore grammatical and structural features.

Latent Sementic Analysis (LSA)\cite{LSA}, which became popular in the 1990s, has opened up a new idea. By mapping utterances to a low-dimensional continuous space of equal length, similarity computation can be performed on this implicit latent semantic space. Since then, more advanced probabilistic models such as PLSA (Probabilistic Latent Semantic Analysis)\cite{PLSA}, which gives a probabilistic interpretation to LSA through a generative model, and LDA (Latent Dirichlet Allocation)\cite{LDA}, which introduces the concept of prior distribution of parameters based on PLSA's model, have been designed. And gradually formed a very hot direction of topic modeling techniques. These techniques provide a concise and convenient semantic representation of text, and make up for the shortcomings of traditional lexical matching methods.

\subsection{Attention Mechanism}
Originally used in machine translation\cite{DBLP:journals/corr/BahdanauCB14}, AM(Attention Mechanism) has become an important concept in the field of neural networks. In the field of artificial intelligence, attention has become an important part of the structure of neural networks and has a large number of applications in natural language processing, statistical learning, speech and computing. Attention mechanism can be explained intuitively using human visual mechanisms. For example, our visual system tends to focus on the part of information in an image that aids in judgment and ignore irrelevant information. Similarly, in problems involving language or vision\cite{DBLP:conf/cvpr/YouJWFL16}, some parts of the input may be more useful for decision making than others\cite{DBLP:conf/ijcai/LiZL17}. For example, in translation\cite{DBLP:journals/corr/BahdanauCB14} and summarization\cite{DBLP:conf/emnlp/RushCW15,DBLP:conf/kbse/WanZYXY0Y18} tasks, only certain words in the input sequence may be relevant for predicting the next word. Similarly, in a code search problem, only certain words in the input description may be more relevant to a particular word of the searched code fragment. The attention mechanism helps to perform the task at hand by allowing the model to dynamically focus on certain parts of the input that help to perform the task at hand.

\subsection{Graph Neural Networks}
Although traditional deep learning methods have been applied to extract features from Euclidean space data with great success, many practical application scenarios in which the data are generated from non-Euclidean space, the performance of traditional deep learning methods on processing non-Euclidean space data is still unsatisfactory. Recent research has focused on the extension of deep learning methods to graphs. Driven by the success of multiple factors, researchers have borrowed ideas from convolutional networks, recurrent networks and deep autoencoders to define and design neural network structures for processing graph data, that is GNN(graph neural networks)\cite{DBLP:journals/spm/BronsteinBLSV17,DBLP:journals/debu/HamiltonYL17}.

Code, as a computer instruction, has similarity to natural language in terms of vocabulary, but is very different from natural language in terms of syntax and extensibility, and is more rigorous than natural language. Understanding the semantics of code through words alone often fails to understand the deeper semantic information expressed in code. Abstract Syntax Tree (AST), as an intermediate representation of high-level language and machine instructions, contains the logical information of the code. Recent researches like SBT\cite{DBLP:conf/iwpc/HuLXLJ18}, Tree-
LSTM\cite{DBLP:conf/acl/TaiSM15}, etc., all serialize the abstract syntax tree, but transforming the topology into serial structure will inevitably lose a lot of information, and it is reasonable to use GNN, which specializes in extracting graph structure features, to extract code structure features.

\section{Proposed Approach}

This section first introduces the whole workflow of code search, and then presents the details of our proposed model architecture for CSSAM.

Figure 1 shows the workflow of code search\cite{DBLP:conf/icse/GuZ018, DBLP:conf/kbse/WanSSXZ0Y19, DBLP:conf/acl/HaldarWXH20}, and our model also follows this process. The code search can be divided into two parts: offline training and online search, where the offline training phase requires us to extract a large number of $<code, decsting>$ pairs from the open source code repository as training data, and then train our model CSSAM with the training data, and after the training is completed, we get a trained search model. Another phase is the online search phase, in which we only need to enter a natural language description and the trained model will search for the closest code fragment.

The overall structure of the CSSAM model is shown in Figure 2, which can be divided into two parts: semantic level matching and structural level matching. In the text semantic level matching module, this paper designs a matching module based on residual interaction to improve the accuracy of semantic matching; in the structure level matching module, this paper designs a new code representation structure of CSRG (Code Semantic Representation Graph) for structural feature extraction; in the similarity calculation module, a fusion attention machine layer to balance the matching contribution of each part.

\subsection{Embedding}
The word embedding layer uses unsupervised learning fasttext\cite{joulin2016fasttext} model, and uses various feature enhancement methods including n-gram and subword regularization\cite{bojanowski2016enriching}. Code and Docstring are tokenized and fed into the word embedding model for word vector training, respectively.

For the graph structure representation of the code, our model needs to learn the structural and semantic information between the local to the global within the code, so as to extract the features embedded in the different structures in multiple structural representations, in order to make each token node able to obtain the local relationships between the internal nodes, we design the graph embedding layer by using Deepwalk\cite{deepwalk} method, and DeepWalk is divided into two parts: random walk and generation of representation vectors. Firstly, some vertex sequences are extracted from the graph using the Randomwalk algorithm; then, with the help of natural language processing ideas, the generated fixed-point sequences are regarded as sentences composed of words, and all the sequences can be regarded as a large corpus, and finally, each vertex is represented as a vector of dimension $d$ using the natural language processing tool word2vec\cite{DBLP:conf/icml/LeM14,w2v}. The flow of the DeepWalk is depicted as Algorithm 1.
\begin{algorithm}[htb]
\caption{DeepWalk$G, w, d, \gamma , t$}
\label{alg:algorithm}
\textbf{Input}: graph $G(V,E)$,
    window size $w$,
    embedding size $d$,
    walks per vertex $\gamma$,
    walk length $t$\\
\textbf{Output}: matrix of vertex representations $\Phi \in R^{V} \times d$.
\begin{algorithmic}[1] 
\STATE Initialization: Sample $\Phi$ from $\mathcal{U}^{V}\times d$
\STATE Build a binary Tree $T$ from $V$
\FOR{i = 0 to $\gamma$}
    \STATE $\mathcal{O}$ = Shuffle($V$)
    \FOR{each $v_{i} \in \mathcal{O}$}
        \STATE $\mathcal{W} = RandomWalk(G,v_{i},t)$
        \STATE SkipGram($\Phi, \mathcal{W}, w$)
    \ENDFOR
\ENDFOR
\end{algorithmic}
\end{algorithm}
\subsection{Code Semantic Representation Graph}
Intermediate representation structures of codes like AST, CFG, DFG, etc. can better reflect the semantic features of codes, however, the increase of model complexity and the improvement of model effectiveness are not proportional when dealing with these tree structures separately. In order to better integrate these code property graphs, we propose a code semantic representation graph based on the abstract syntax tree.

For snippets with a very large amount of code, there is the problem of long coding spans when data or variables are called repeatedly, for example, a function customizes a new variable in the first several lines but does not call it again until the code's last line of the function; consequently, when tokens features are extracted by models such as GNN or RNN, the long-term relationship of that variable in the code text would disappear due to the increase in sequence length. Hence, the data dependency of each statement can be determined by analyzing the data stream that the statement uses. In order for the semantic representation model to be able to compensate for the inherent flaws in the feature extraction algorithm (such as the model's inability to capture long-range dependencies, etc.), the representation model we designed incorporate the acquisition of contextual information of code statements, by adding directed edges to variables in the context, the model could eliminate the problem of disappearing code semantics.

To minimize information loss and maintain the original AST structure, this paper proposes CSRG (Code Semantic Representation Graph), a graph structure based on AST that aggregates node information. CSRG is based on the abstract syntax tree but more compact, and keeps the data flow information in the CSRG structure.

\begin{figure}[htb]
  \centering
  \includegraphics[width=\linewidth]{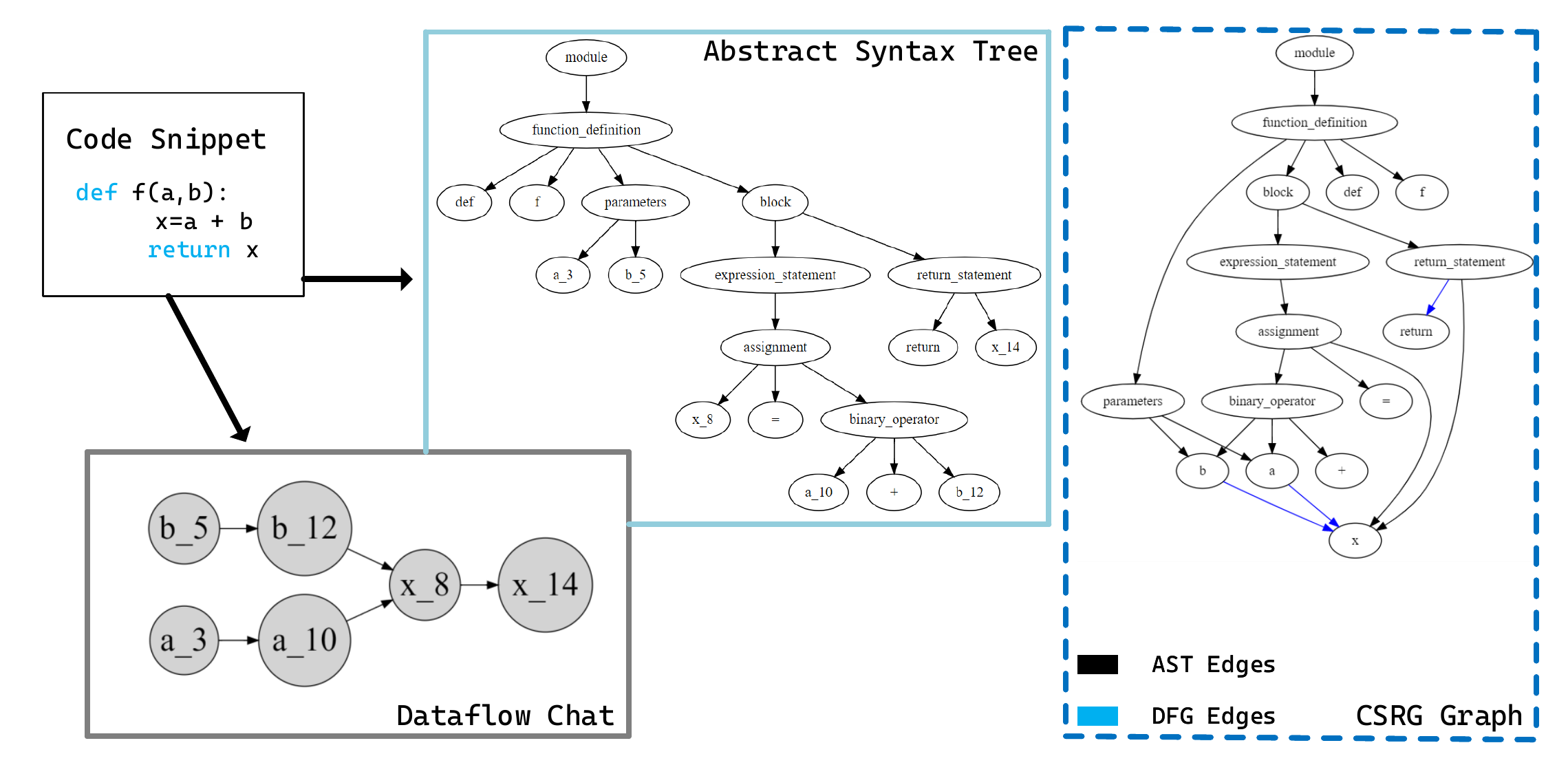}
  \caption{Example of code semantic representation graph. There are two types of edges in CSRG, one from AST and the other from DFG.}
\end{figure}
Figure 3 shows the generation process of the code semantic representation graph. We generate the abstract syntax tree(AST) and data flow graph(DFG) of the code segment by third-party tools\footnote{https://tree-sitter.github.io/tree-sitter/}. 

We first generate the DFG abstract representation of the code using the tree-sitt tool, and then use depth first search algorithm to locate the corresponding DFG node in the AST. We can notice that there are many nodes in the AST and DFG, Nodes represent variables such as a\_\{$i$\}, b\_\{$i$\} and x\_\{$i$\}, where $i$ indicates its position in the sequence of tokens of the code. Nodes in the DFG can be located in the AST. To obtain CSRG, we firstly de-duplicate the nodes with the same attributes, ignore the indicates. Then we get the AST and DFG with reduced number of nodes. Secondly, find the corresponding DFG node in the AST, and add the node relationship from the DFG to the AST. The improved AST now contains two types of edges, one that is present in the AST itself and another that is added based on the DFG. We can assign different weights to these two types of edges. As a result of the rule for generating three-address codes, variables and attributes are typically located at the leaf node positions in the AST. After combining the data flow in this structure, directed edges are added between each leaf node to determine the relationship between branch statements in the AST, which further reduces the sequence span during feature extraction. Finally, the strings in the nodes are replaced based on the syntax parsing dictionary of the source code in order to obtain the complete semantic flow graph of the code. And now we obtained the code semantic representation graph, which has a smaller number of nodes compared to AST, and also has data flow feature.

\subsection{Context Representation}
For natural language descriptions, current feature extraction models have achieved very good results in the field of natural language processing. For long texts, Transformer can be selected for feature extraction, and since the length of code description statements is generally short, our model uses LSTM\cite{DBLP:conf/nips/HochreiterS96} for the extraction of semantic features of descriptions.
\subsection{Graph Representation}
Code Semantic Representation Graph CSRG as a graph structure, using GNN to encode can dig deeper into its node features and patterns.  GCN(Graph Convolutional Network)\cite{DBLP:conf/iclr/KipfW17} can perform convolutional operations on Graph. However, GCN has some drawbacks: it relies on Laplacian matrix and cannot be used directly on directed graphs; model training relies on the whole graph structure and cannot be used on dynamic graphs; different weights cannot be assigned to neighboring nodes during convolution processing. Therefore Graph Attention Network \cite{DBLP:conf/iclr/VelickovicCCRLB18} GAT (Graph Attention Network) is proposed to solve the problems of GCN, and we use GAT to extract CSRG features of the code.

Suppose Graph contains $N$ nodes, each node has a feature vector of $h_{i}$ and dimension $F$, denoted as $h = \{ h_{1}, h_{2}, ... , h_{N}\}, h_{i} \in \mathbb{R}^{F}$, a linear transformation of the node feature vector $h$.
\begin{equation}
    h_{i}^{'} = Wh_{i} ,W \in \mathbb{R}^{F^{'}\times F}
\end{equation}
\begin{equation}
    h_{'} = \{h^{'}_{1}, h^{'}_{2}, ..., h^{'}_{N}\}, h^{'}_{i} \in \mathbb{R}^{F^{'}}
\end{equation}
where $h_{i}^{'}$ is the new eigenvector, $F^{'}$ denotes the dimension, and $W$ is the linearly transformed matrix.

If node $j$ is a neighbor of node $i$, the importance of node $j$ to node $i$ can be calculated using the $Attention$ mechanism, the $Attention Score$.
\begin{equation}
    e_{i j}=\text { Attention }\left(W h_{i}, W h_{j}\right)
\end{equation}
\begin{equation}
    \alpha_{i j}=\operatorname{Softmax}_{j}\left(e_{i j}\right)=\frac{\exp \left(e_{i j}\right)}{\sum_{k \in N_{i}} \exp \left(e_{i k}\right)}
\end{equation}

The specific $Attention$ approach of GAT is as follows: the feature vectors $h_{i}^{'}, h_{j}^{'}$ of nodes $i,j$ are spliced together and the inner product is calculated with a $2F^{'}$ dimensional vector $a$. The activation function uses $LeakyReLU$ with the following equation:
\begin{equation}
    \alpha_{i j}=\frac{\exp \left(\text { LeakyReLU }\left(a^{T}\left[W h_{i} \| W h_{j}\right]\right)\right)}{\sum_{k \in N_{i}} \exp \left(\text { LeakyReLU }\left(a^{T}\left[W h_{i} \| W h_{k}\right]\right)\right)}
\end{equation}
where $\|$ indicates a splicing operation.

The feature vector of node $i$ after $Attention$ is as follows:
\begin{equation}
    h_{i}^{\prime}=\sigma\left(\sum_{j \in N_{i}} \alpha_{i j} W h_{j}\right)
\end{equation}
where $\sigma$ represents a nonlinear activation function.
\subsection{CRESS Block(Cross And Residual)}
Since the lexical and syntactic structures of code and natural language are different, it is a great challenge to align code and text across sequences to keep the vectors spatially consistent during similarity matching. This is necessary for accurate semantic relationship between code and query. Our task is essentially to find a kind of mapping from code snippet to corresponding description, that is to find the matching relationship between code snippet and description, and how to accurately learn the deeper matching relationship between them is a major difficulty in the code search task\cite{DCNV2}. At the lexical level, we can consider code and description as two different languages, and in order to better match two kinds of texts just from words and sentences, this paper proposes a matching module based on residuals and interactions, called CRESS Block. The CRESS block uses bidirectional version of residual connections to cross consecutive query text and code text, it could provide more features for alignment procedures. Specifically, the sequence encoder first computes the contextual features of the query text and code text sequences within the CRESS block. The encoder's inputs and outputs are then concatenated and fed to the alignment layer, which uses an enhanced residual concatenation to model the alignment and interaction of the two sequences. A fusion layer unites the alignment layer's inputs and outputs. Finally the output of the code and query are then sent to the pooling layer, converted to a vector of fixed length, and used as input for similarity matching. implemented as shown in Figure 4.
\begin{figure}[H]
    \centering
    \includegraphics[width=\linewidth]{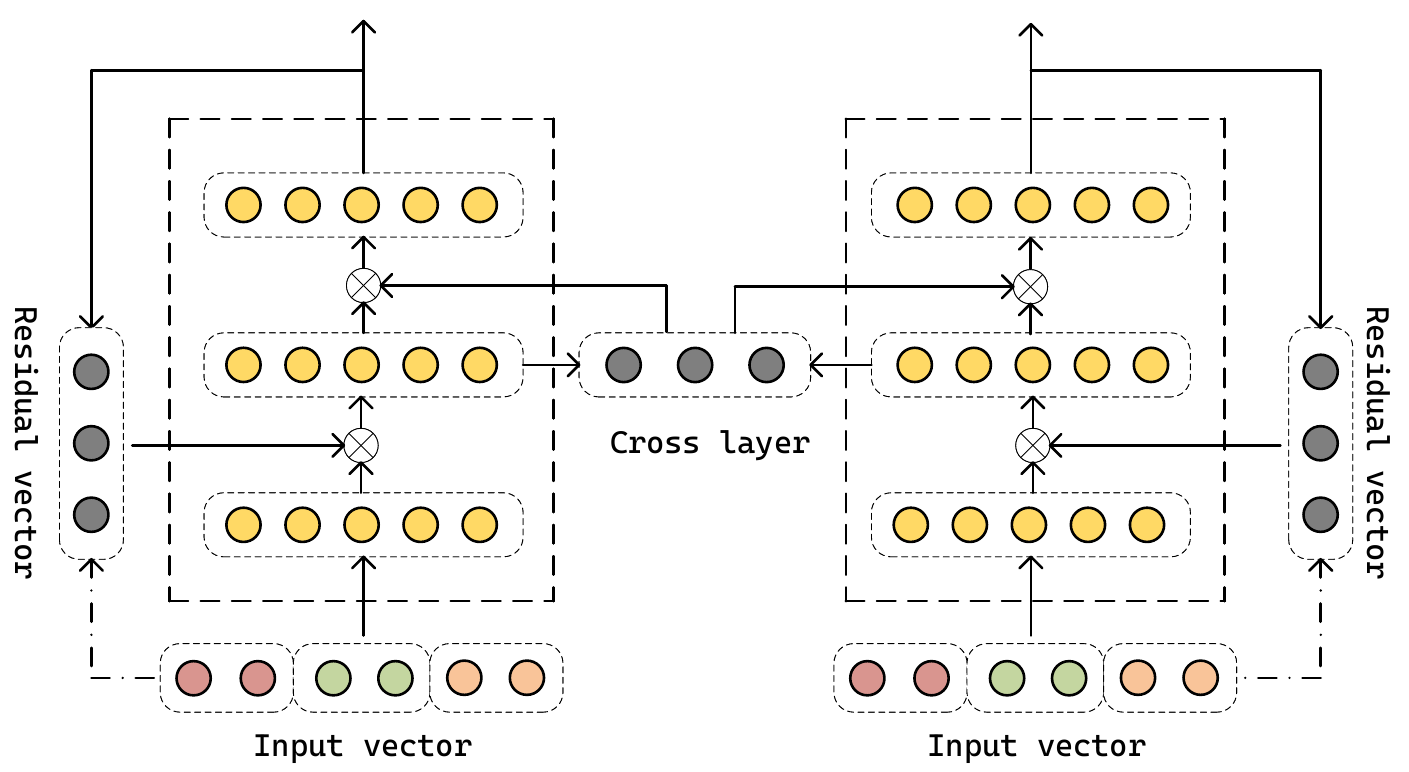}
    \caption{Details of a CRESS block. The operation $\mathrm{x}_{l+1}=\mathrm{x}_{0} \odot\left(W_{l} \mathrm{x}_{l}+\mathrm{b}_{l}\right)+\mathrm{x}_{l}$ is denoted as $ \otimes $.}
    
\end{figure}

For a sequence of length $l$, we denote the input and output of the $n$'s block as $x^{(n)} = (x^{(n)}_{1},x^{(n)}_{2},...,x^{(n)}_{l})$ and $o^{(n)} = (o^{(n)}_{1},o^{(n)}_{2},...,o^{(n)}_{l})$ the input of the $n$'s block is the concat of the first input and the output of previous two blocks.
\begin{equation}
    x^{(n)}_i = [x^{(1)}_{i};o^{(n-1)}_{i}+o^{(n-2)}_{i}]
\end{equation}
where [;] denotes the concat operation, which facilitates the subsequent matching with the addition of residual information.

The cross layer based on the attention mechanism, takes features from the two sequences as input and computes the aligned representations as output. Input from the first sequence of length $l_{a}$ is denoted as $a = (a_{1}; a_{2};...; a_{l_{a}})$ and input from the second sequence of length lb is denoted as $b = (b_{1}; b_{2};...;b_{l_{b}})$. The similarity score $e_{ij}$ between $a_{i}$ and $b_{j}$ is computed as the dot product of the projected vectors:
\begin{equation}
    e_{ij} = F(a_{i})^{T}F(b_{j})
\end{equation}
$F$ is an identity function or a single-layer feedforward network. The choice is treated as a hyperparameter.The output vectors $a'$ and $b'$ are computed by the similarity score $e_{ij}$:
\begin{equation}
     a^{'}_{i} = \sum_{j=1}^{l_{b}}\frac{exp(e_{ij})}{{\textstyle \sum_{k=1}^{l_{b}}} exp(e_{ik})}b_{j}
\end{equation}
\begin{equation}
     b^{'}_{j} = \sum_{i=1}^{l_{a}}\frac{exp(e_{ij})}{{\textstyle \sum_{k=1}^{l_{a}}} exp(e_{kj})}a_{i}
\end{equation}

The residual vector is the output of the previous block. For the first block, its input and residual vectors are the original embedding vectors in Figure 2(docstring vector, tokens vector).
\subsection{Attention Fusion}
By introducing the above modules, for the code we generate two vectors, $CodeTokens_{CRESS}$ and $CSRG_{GAT}$, and for their corresponding descriptions we generate two vectors, $DocTokens_{lstm}$ and $DocTokens_{CRESS}$ , and we need to merge these corresponding vectors are fused into one vector, and here we use the attention mechanism. In the text sequence model, the importance of each word to the final matching result is different, so we need to calculate the attention score for each word and obtain the final vector by weighting.
\begin{equation}
    v=\sum_{i=1}^{T} \alpha_{i} h_{i}
\end{equation}
where $\alpha_{i}$ denotes the attention score corresponding to the hidden state $h_{i}$ of the encoder.

Now that we have the vectors obtained by the different modules and the attention scores corresponding to each vector, we can obtain the code and describe the respective vectors.
\begin{equation}
    CodeTokens_{CRESS}  = \alpha^{t o k_{RE2}} \mathbf{h}^{t o k_{RE2}}
\end{equation}
\begin{equation}
    CSRG_{GAT}  = \alpha^{CSRG} \mathbf{h}^{CSRG}
\end{equation}
\begin{equation}
    Docs_{lstm} = \alpha^{t o k_{lstm}} \mathbf{h}^{t o k_{lstm}}
\end{equation}
\begin{equation}
    Docs_{CRESS}  = \alpha^{t o k_{CRESS}} \mathbf{h}^{t o k_{CRESS}}
\end{equation}
\begin{equation}
    \mathbf{x}_{code}=[CodeTokens_{CRESS}; CSRG_{GAT}]
\end{equation}
\begin{equation}
    \mathbf{x}_{docs}=[DocsTokens_{lstm} ; DocsTokens_{CRESS}]
\end{equation}
where$[;]$denotes the concat operation.
\subsection{Loss Function}
We have described above how to obtain the vectors corresponding to the code and the textual description. And we already know that semantically more similar statements than embedding them into the same high-dimensional vector space, the corresponding vector distances obtained are also more similar. In other words, given a code fragment $x$ and a descriptive statement $d$, if they are corresponding, then the vector similarity of our model embedding has the highest degree of similarity, and if they are not in correspondence, then the vectors embedded in the model should have the smallest degree of similarity. In the training phase, we use the triple $<x,d+,d->$ to train the model, where $d+$ denotes the description corresponding to the code segment $x$, and $d-$ denotes the description not related to the code segment, and we want the model to maximize the similarity between $<x,d+>$ and minimize the similarity between $<x,d->$. In summary, the model's loss function is shown as follows:
\begin{small}
    \begin{equation} 
        \mathcal{L}(\theta)=\sum_{<x, d^{+},d^{-}>\in\mathcal{D}}\max\left(0,\beta-\operatorname{sim}\left(\mathrm{x},\mathrm{d}^{+}\right)+\operatorname{sim}\left(\mathrm{x}, \mathrm{d}^{-}\right)\right)
    \end{equation}
\end{small}
where $\theta$ denotes the model parameters, $\mathcal{D}$ denotes the training data set, $\beta$ is a hyperparameter, $sim$ denotes the similarity score between two vectors, $\mathrm{x},\mathrm{d}^{+},\mathrm{d}^{-}$ denotes the code fragment $x$, the description statement $d+$ corresponding to the code $\mathrm{x}$,and description statements $d-$ not corresponding to code fragments with the same dimension after model embedding.
\subsection{Online Code Search}
Given a set $\chi$ of code segments to be searched, for an input query $q$, the model needs to sort the similarity of all code segments in the database and select the set $x_{1},x_{2},..., x_{k}$ of $k$ code segments that are closest to the query $q$. For the input query $q$, the similarity between $q$ and $x$ is calculated by cosine similarity for each code segment $x$ in the set of code snippets as follows:

\begin{equation}
    \operatorname{sim}(x, q)=\cos (\mathbf{x}, \mathbf{q})=\frac{\mathbf{x}^{T} \mathbf{q}}{\|\mathbf{x}\|\|\mathbf{q}\|}
\end{equation}
where \textbf{x} and \textbf{q} denote the vectors of code segments and query statements, respectively. The larger the value of similarity, the higher the relevance of the corresponding code segment and query statement.
\section{Experiments}
\subsection{General Settings}
To train our proposed model, we first randomize the training data and set the mini-batch size to 32. For each batch, the code is padded with a special token $<PAD>$ to the maximum length. All tokens in our dataset are converted to lower case. We set the word embedding size to 300. For LSTM unit, we set the hidden size to be 256. For CRESS, set 4 blocks of iteration. The margin $\beta$ is set to 0.05. We update the parameters via Adam\cite{DBLP:journals/corr/KingmaB14} optimizer with the learning rate 0.0001. To prevent over-fitting, we use dropout with 0.2. All the models in this paper are trained for 100 epochs. All the experiments are implemented using the PyTorch 1.2 framework with Python 3.8, and the experiments were conducted on HPC with four Nvidia Tesla 100 GPU with 16 GB memory, running CentOS 7.5.
\subsection{Baselines}
We have selected the following models to compare with ours.
\begin{itemize}
    \item \textbf{CodeHow}\cite{2015CodeHow}: It is a classical code search engine proposed in the previous years. It is based on information retrieval code search tool that contains an extended Boolean model and API matching.
    \item \textbf{DeepCS}\cite{DBLP:conf/icse/GuZ018}: A code retrieval method based on deep neural networks. By embedding the source code and description into the same vector space to perform the code-description matching search.
    \item \textbf{MPCAT}\cite{DBLP:conf/acl/HaldarWXH20}: a code search model using hierarchical traversal method to encode code abstract syntax trees and incorporating text matching model BiMPM model.
    \item \textbf{TabCS}\cite{tabcs}: A code search model using two-stage attention for feature extraction, while using association matrix for parameter sharing, which achieves good results.
\end{itemize}

\subsection{Datasets}
We train and evaluate the model on two publicly available datasets. One is the Hu’s \cite{hudataset} dataset\footnote{https://github.com/xing-hu/EMSE-DeepCom} that contains 480k code-query pairs. Hu’s dataset was collected from GitHub’s Java repositories created from 2015 to 2016. To filter out low-quality projects, Hu et al. \cite{hudataset} only considered the projects with more than ten stars. Then, they extracted Java methods and their corresponding Java doc from these Java
projects. The first sentence of the Javadoc is considered as the query. The other is the Husain’s \cite{CSN} dataset\footnote{https://github.com/github/CodeSearchNet}. The corpus contains 540k Java code with queries written in natural language collected from GitHub repositories. We filter out the code snippets following this criterion: the first sentence of its annotation should be longer than two words. After filtering, we obtain a train set containing 330k annotation-function pairs and a test set containing 19k
annotation-function pairs. The statistics of the two datasets are shown in Table 1.
\begin{table}[h]
\centering
\caption{Datasets Details}
\begin{tabular}{c|c|c}
\hline
 Dataset & Hu’s dataset & Husain’s dataset\\ 
 \hline
Train & 475812   & 329967         \\ 
Test & 10000   & 19015          \\ 
Avg. tokens in comment & 10.25   & 10.2          \\ 
Avg. tokens in code & 58.6   & 68.5          \\
Max tokens in comment & 32  & 440 \\
Max tokens in code & 841  & 369\\
 \hline
\end{tabular}
\end{table}

\begin{table*}[htb]
\centering
\caption{Ablation Study}
\begin{tabular}{l|c|c|c|c|c}

\hline
Model & SR@1 & SR@5 & SR@10 & MRR & NDCG@50 \\ 
\hline
Base & 0.1426   & 0.2901   & 0.4147    & 0.2313   & 0.3065 \\ 
Base+CSRG & 0.1982   & 0.3390   & 0.4865    & 0.2553   & 0.3267       \\ 
Base+CRESS & 0.2286   & 0.4642   & 0.5735    & 0.3407    & 0.4327 \\ 
Base+CRESS+CSRG & 0.2611   & 0.4893   & 0.6122    & 0.3854   & 0.4797       \\ 
Base+CRESS+CSRG+Attn & \textbf{0.3221}   & \textbf{0.6362}   & \textbf{0.7051}    & \textbf{0.4831}   & \textbf{0.5843}      \\ 
\hline
\end{tabular}
\end{table*}
\subsection{Evaluation Metrics}
\subsubsection{\emph{MRR(Mean Reciprocal Rank)}}
This is a commonly used metric to measure the effectiveness of search algorithms, and is now widely used in problems that allow multiple results to be returned, or problems that are currently difficult to solve (since the accuracy or recall would be poor if only $top1$ results were returned, multiple results are returned first in cases where the technology is not mature). In such problems, the system gives a confidence level (score) to each returned result, and then sorts the results according to the confidence level, returning the results with high scores first.
	
For a query set $Q$, the set of returned results is $q$, the correct result appears at $FRank$, and the score is the reciprocal of $FRank$, then $MRR$ is:
\begin{equation}
    M R R=\frac{1}{|Q|} \sum_{q=1}^{|Q|} \frac{1}{F R a n k_{q}}
\end{equation}
Higher MRR values indicate better performance of the code search model.
\subsubsection{\emph{SuccessRate@k(Success Percentage at k)}}
This metric measures the percentage of the first $k$ results for which one or more correct results may exist, and is calculated as follows:
\begin{equation}
    \text { SuccessRate@k }=\frac{1}{|Q|} \sum_{q=1}^{Q}\delta\left(F \text { Rank }_{q} \leq k\right)
\end{equation}
where $\delta $ is a function, the output is 1 when the input is true, otherwise the output should be 0. A good code search engine should place the correct results as close to the front of the return value as possible, so that users can easily find the results they need more quickly, and similarly, the higher the R@k value, the better the performance of the code search model.
\subsubsection{\emph{Normalized Discounted Cumulative Gain(NDCG)}}
The normalized discounted cumulative gain is used as an evaluation metric for the sorting results to evaluate the accuracy of the sorting.
Recommender systems usually return a $item$ list for a user, and assuming the list length is $K$, the difference between this sorted list and the user's real interaction list can be evaluated with $NDCG@K$.
\begin{equation}
    NDCG@k = \sum_{i=1}^{k} \frac{2^{r(i)}-1 }{log_{2}(i+1) }
\end{equation}
where $r(i)$ is the score of the ith result. In the code search task, only correct or incorrect, the corresponding scores are 1 and 0. In our experiments, $NDCG@50$ is taken as the evaluation index considering the dataset size.
\subsection{Implementation Details}
\subsubsection{Tokenization}
As a natural language, it is enough to separate words according to their spacing, but for programming languages, there are elements such as hump nomenclature and a lot of symbolic language. For hump nomenclature, such as ``getFileNam", we can split it into three words: ``get", ``file" and ``name". For the large number of symbols in the programming language, in some papers all the symbols are removed, leaving only the words, but in fact the symbols in the code also contain a lot of semantic information, so in this paper, we keep the symbols in the code syntax

\subsubsection{Bias of Code Semantic Representation Graph}
To distinguish between the two types of edge information in the code semantic graph. After a variety of weight settings, we found that the weight setting of the edge in CSRG has little effect on the accuracy of the search results. We set the weight of the edge generated by AST to 0.4, the weight of the DFG edges is set to 0.6, can achieve a relatively better result. 

\section{RESULTS}
To evaluate our proposed approach, in this section, we conduct experiments to answer the following questions:
\begin{itemize} 
\item RQ1. Does our proposed approach improve the performance of code retrieval when compared with state-of-the-art approaches?
\item RQ2. What is the effectiveness and the contribution of each strategy e.g., CRESS layer, CSRG of source code for the final retrieval performance, and what about their combinations?
\item RQ3. What's better about using CSRG than using AST or DFG?
\item RQ4. What is the performance of our proposed model when varying the CRESS block number, code CSRG node number?
\item RQ5. Is the search result of CSSAM better than state-of-the-art approaches?
\end{itemize}

We ask RQ1 to evaluate our deep learning-based model compared to some state-of-the-art baselines, which will be described in the following subsection. We ask RQ2 in order to evaluate the performance of each module. We ask RQ3 to analyze the performance of our proposed CSRG. We ask RQ4 to analyze the sensitivity of our proposed model when varying the CRESS block number, code CSRG node number. We ask RQ5 to verify the searching results of our proposed CSSAM. 

\subsection{RQ1: Does our proposed approach improve the performance of code retrieval when compared with state-of-the-art approaches?}

We trained and tested CodeHow, DeepCS, TabCS, MPCAT, and our model CSSAM on the same dataset, and the experimental data are shown in Tables 3 and 4. From the data, our model outperforms three deep learning-based models (DeepCS, TabCS, MPCAT) and one information retrieval-based model codeHow.

For Hu et al.’s dataset, the results in Table 3, our model achieves an MRR value of 0.483  and 0.322/0.636/0.705 in SR@/1/5/10, CSSAM outperforms CodeHow, DeepCS, TabCS, and MPCAT 80.39\%, 70.04\%, 4.82\%, 2.33\% in MRR, respectively. 
For Husain’s dataset, the results in Table 4, our model achieves an MRR value of 0.394 and 0.259/0.491/0.575 in SR@/1/5/10, CSSAM outperforms CodeHow, DeepCS, TabCS, and MPCAT in MRR and SR@k.

The experimental results show that our model performs better than the above baseline model.

\subsection{RQ2: What is the effectiveness and the contribution of each strategy e.g., CRESS layer, CSRG of source code for the final retrieval performance, and what about their combinations?}
We did five sets of ablation experiments based on the model to verify the effect of each module on the experimental results. Table 2 shows the effect of each module on the experimental results. From the experimental data, we can see that both the semantic level matching module and the structural level matching module have positive effects on the experimental results, and the model performs better after fusing the matching modules of each level than using these modules alone, which also indicates that the complementary effects between the matching layers at different levels outweigh the conflicts between them. The experimental results also demonstrate the positive contribution of fused attention layers to the model effects by adding and removing Attention layers.
\begin{table}[htb]
\centering
\caption{Experiments on Hu et al.’s dataset.}
\scalebox{0.85}{
\begin{tabular}{c|c|c|c|c|c}
\hline
Model & SR@1 & SR@5 & SR@10 & MRR & NDCG@50 \\ 
\hline
CodeHow & 0.2341   & 0.3585   &0.4944     & 0.2678   & 0.3411       \\ 
\hline
DeepCS & 0.2574   & 0.3897   & 0.5169    & 0.2834   & 0.3690       \\ 
\hline
 MPCAT & 0.2997   & 0.4946   & 0.6642    & 0.4721   & 0.4891       \\ 
\hline
TabCS & 0.3152   & 0.5744   & 0.6831    & 0.4609   & 0.5492       \\ 
\hline
CSSAM & \textbf{0.3221}   & \textbf{0.6362}   & \textbf{0.7051}    & \textbf{0.4831}   & \textbf{0.5843}       \\ 
\hline
\end{tabular}
}
\end{table}

\begin{table}[htb]
\centering
\caption{Experiments on Husain’s dataset}
\scalebox{0.85}{
\begin{tabular}{c|c|c|c|c|c}
\hline
Model & SR@1 & SR@5 & SR@10 & MRR & NDCG@50 \\ 
\hline
CodeHow & 0.1671   & 0.3947   & 0.4420    & 0.2459   & 0.2901     \\ 
\hline
DeepCS & 0.1460   &0.3562   & 0.4036    & 0.2243   & 0.2617       \\ 
\hline
 MPCAT & 0.1711   & 0.3776   & 0.4414    & 0.2936   & 0.3496       \\ 
\hline
TabCS & 0.2259   & 0.4195   & 0.4910    & 0.3487   & 0.3873      \\ 
\hline
CSSAM & \textbf{0.2585}   & \textbf{0.4912}   & \textbf{0.5754}    & \textbf{0.3941}   & \textbf{0.4223}       \\ 
\hline
\end{tabular}
}
\end{table}

As we can see in table 2, each strategy has a positive impact on search results. CSRG brings structure information to the model, and the code features are more complete. CRESS matches the code and description from the word level, which increases the performance of the model at the fine-grained level. " "Unity is strength", which is more obvious under the blessing of the attention mechanism.
\subsection{RQ3: What's better about using CSRG than using AST or DFG?}
In order to explore why CSRG is effective, we conducted performance tests according to different graph structures. The results are shown in Table 5. We can find that using graph structure information can significantly improve the code search results, but it will also bring the burden of training. Compared with only using AST, our proposed CSRG significantly reduces the complexity of input data and improves the training speed. Compared with just using DFG, our CSRG significantly improves the accuracy of search results.
\begin{table}[ht]
\centering
\caption{Training Details}
\begin{tabular}{l|c|c}
\hline
 Model & Training time & MRR Score\\ 
 \hline
CSSAM-{w/o.Graph}& 16.2 hours   & 0.292         \\ 
CSSAM-{w.AST} & 27.1 hours  & 0.337          \\ 
CSSAM-{w.DFG} & 18.0 hours   & 0.304          \\ 
CSSAM-{w.CSRG} & 24.8 hours   & 0.348          \\
 \hline
\end{tabular}
\end{table}

There is no doubt that structural information is very useful for code search. After adding AST and DFG information, the code search MRR is improved. We all know that ast has more node information than DFG, which also leads to the training time of AST model is longer than DFG model. And using CSRG, although the training time is increased compared with baseline, it is less than that of AST. At the same time, MRR is also higher than that of AST and DFG models.
\subsection{RQ4: What is the performance of our proposed model when varying the CRESS block number, code CSRG node number?}
In order to explore the robustness of this model, we analyzed two hyperparameters that may affect the effect of the model, which are number of CSRG nodes. Figure 5 shows the effects of each parameter on the model evaluation metrics, and firstly, we can see that our model performs relatively stable when using different hyperparameters, the robustness of our model can be proved.
\begin{figure}[htb]
    \centering
    \includegraphics[width=\linewidth]{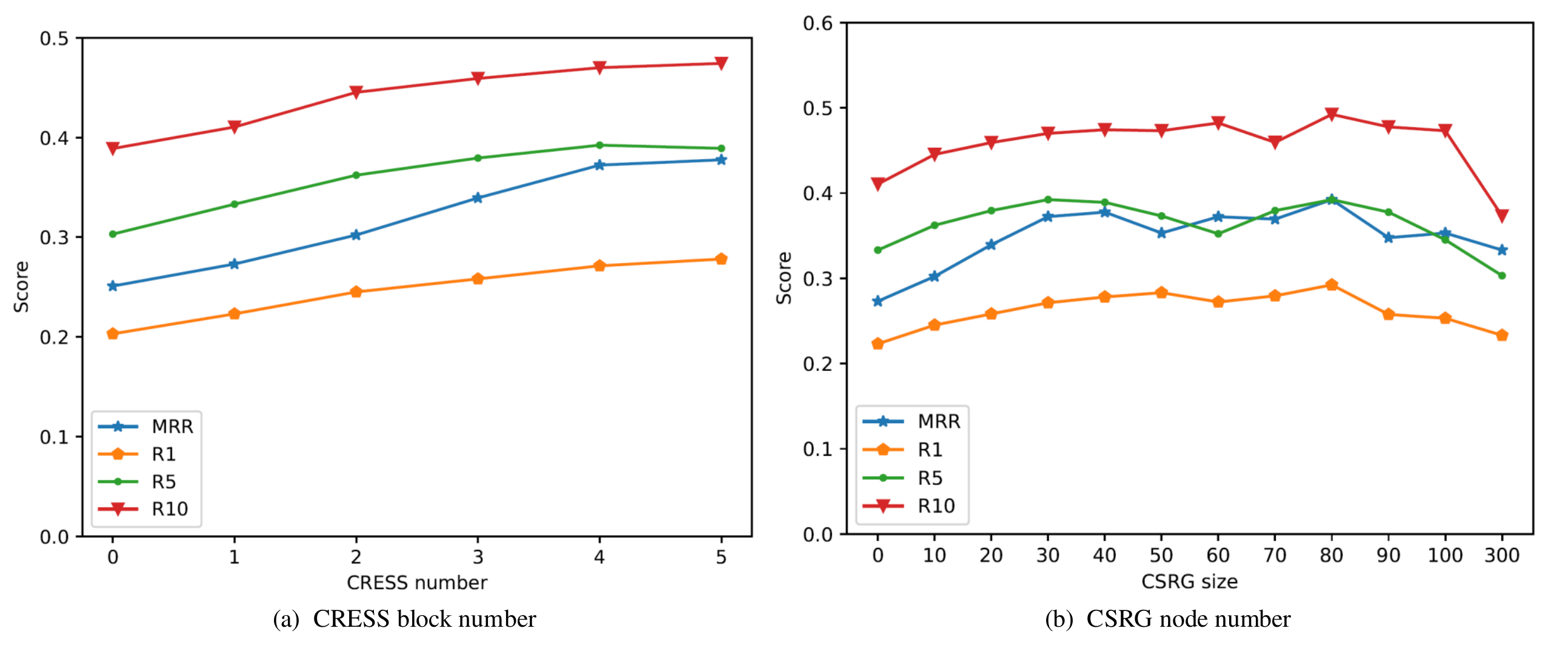}
    \caption{Sensitivity Analysis}
\end{figure}

We have tried the number of CRESS blocks from 0 to 5. It can be found that with the increase of the number of layers, all the indicators of the model show an upward trend, but followed by a significant increase in the consumption of the model. We find that when the number of cress is set to 5, the growth of the model indicators has slowed down. Taking a comprehensive consideration, we take 4 as the optimal parameter.

For the number of CSRG nodes, at first, as the number of nodes increases, the indicators are in a stable growth trend, but when the number of nodes exceeds 80, the indicators appear to decline, and finally we tried the case of 300 nodes, the indicators are still not as good as the optimal point, for this phenomenon, the reason of occurrence is related to the dataset, when the number of nodes is close to the average of this dataset, the model can be made optimal, too few nodes cannot provide enough useful information, too many nodes means more padding, which is equivalent to diluting the information and cannot achieve the optimal performance of the model.

\subsection{RQ5: Is the search result of CSSAM better than state-of-the-art approaches?}
Figure 6 shows the first retrieved results of CSSAM and other models, for queries ``save string into the file” and``converts the given hex string into a plain string". We can notice the CSSAM returns the correct code snippet.But neither DeepCS nor TabCS returned the correct results, they just matched the correlation between the code and the description at the tokens level. The fusion of DFG and AST enriches the semantic and associative information present in the nodes inside the code snippet, and the CRESS block we proposed further aligns the textual features of the code and query semantically, thus allowing our model to learn additional contextual features based on the sequence structure. On the other hand, CSSAM not only matches at the tokens level, but also takes into account the deep semantic information of the code, so the search results are more accurate. 
\begin{figure}[htb]
    \centering
    \includegraphics[width=\linewidth]{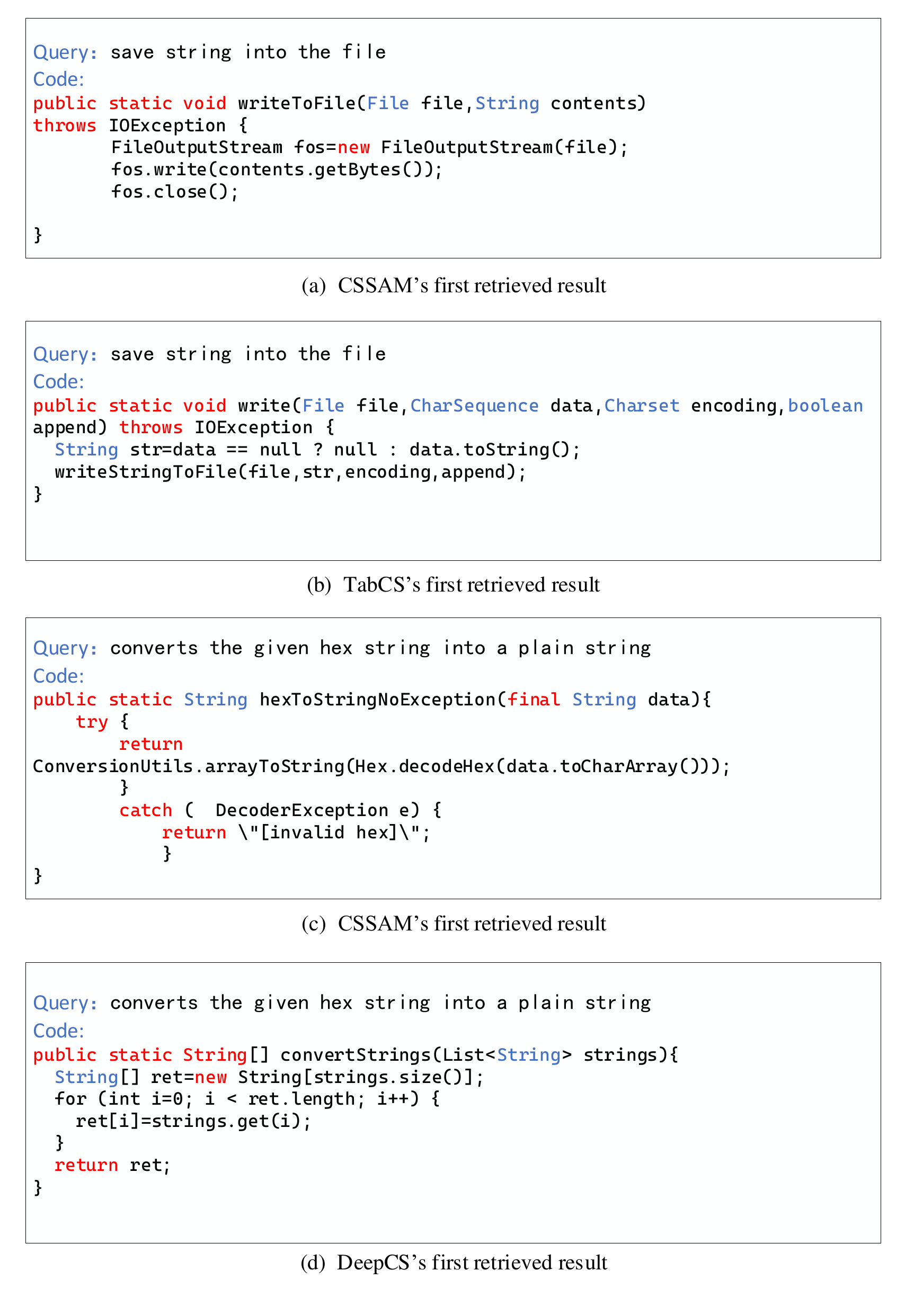}
    \caption{First retrieved results for different models}
\end{figure}

\section{Discussion}
\subsection{Strength of CSSAM}
We have identified three advantages of CSSAM, which may explain its effectiveness in code retrieval: (a) a more comprehensive representation of source code. CSSAM uses CSRG, which contains supplementary information in addition to the code tokens representation. (b) A fine-grained matching module cress, we have proved the effectiveness of this module through experiments. And (c) a learning the unified framework source code of heterogeneous representation and the intermediate semantic space of description. CSSAM is an end-to-end neural network model, which uses a unified architecture to learn the source code to represent the code and description in the intermediate semantic space.

\subsection{Threats to Validity}
Our proposed CSSAM may suffer from two threats to validity and limitations. One threat to validity is on the evaluation
metrics. In practice, there are multiple related code snippets for a query. However, for automated evaluation, each query has limited correct code snippets. During the evaluation, the other results, which are also related, can not be recognized unless human involves. 

Another threat to validity lies in the extensibility of our proposed approach. Our model has only been trained and tested on Java data sets. For other programming languages, the model results are unknown. At the same time, because CSSAM needs to use CSRG, but this graph structure needs to be generated by our own tools. At present, it is difficult to extract CSRG from any language, which is also the focus of our future work.

\section{Conclusion and Outlook}
In this paper, we propose a code search model CSSAM based on dual-level matching of semantics and structure, which not only considers the matching relationship between code and description at the semantic level, but also proposes CSRG to match code and description at the structural level based on code AST, and adds an attention mechanism to balance the matching results at each level, and experiments prove that our model is effective and achieves state-of-the-art, which exceeds the existing code search models.

In our future work, we plan to continue to improve the model's effectiveness, and hope to achieve language-independence, as the current model performs differently on different language datasets, and the extraction and generation of CSRG cannot be done for all languages. At the same time, our work may also have contributions in other areas related to code such as code generation and code summarization, and this aspect is also worthy of our research.

\bibliographystyle{IEEEtran} 
\bibliography{conference_101719}

\end{document}